\documentclass[12pt]{article}

\usepackage{fancyhdr}
\usepackage{epsfig}
\usepackage{cite}
\usepackage{graphicx}
\usepackage[centertags]{amsmath}
\usepackage{theorem}

\usepackage{graphicx}
\usepackage{amssymb}
\usepackage{latexsym}
\usepackage{epic}
\usepackage{pstricks}
\usepackage{color}
\usepackage{mathrsfs}
\usepackage{latexsym}
\usepackage{epic}
\usepackage{braket}
\usepackage{slashed}
\newcommand{\be}{\begin{equation}}
\newcommand{\ee}{\end{equation}}
\newcommand{\bea}{\begin{eqnarray}}

\newcommand{\eea}{\end{eqnarray}}
\newcommand{\ti}{\widetilde}

\numberwithin{equation}{section}
\def\hybrid{\topmargin 22pt    \oddsidemargin 0pt %%%%%%%%%%%%%% Archive-30pt
        \headheight 0pt \headsep 0pt
        \textwidth 6.25in       % A4 paper
        \textheight 9.0in       % A4 paper
        \marginparwidth .875in
        \parskip 5pt plus 1pt   \jot = 1.5ex}

\hybrid
\usepackage{caption} 
\newcommand{\Tr}{{\rm Tr}}

\newcommand{\gl}{\lambda}
\newcommand{\refe}[1]{Eqn.~(\ref{#1})}

\renewcommand{\thefootnote}{\fnsymbol{footnote}}
\begin{document}
\center{
\begin{flushright} QMUL-PH-10-13
\end{flushright}
\vspace{1cm}
\begin{center}
{\Large\textbf{Warped General Gauge Mediation}}
\vspace{1cm}

\textbf{Moritz McGarrie$^{1,}$\footnote[2]{\texttt{m.mcgarrie@qmul.ac.uk}} and Daniel C. Thompson$^{2,}$\footnote[3]{\texttt{dthompson@tena4.vub.ac.be}} }\\
\end{center}

{\it{ ${}^1$ Queen Mary University of London\\
Center for Research in String Theory\\
Department of Physics\\
Mile End Road, London, E1 4NS, UK.}}

\vspace{0.5 cm}
{\it {${}^2$ Theoretische Natuurkunde, Vrije Universiteit Brussel, and \\
International Solvay Institutes\\
Pleinlann 2, B-1050, Brussels, Belgium}}

%%%%%%%%%%%%%%%%%%%%%%%%%%%%%%%%%%%%%%%%  Resets the counter to 0 and then uses Arabic numerals i.e. 1,2,3...
\setcounter{footnote}{0}
\renewcommand{\thefootnote}{\arabic{footnote}}
%%%%%%%%%%%%%%%%%%%%%%%%%%%%%%%%%%%%%%%%%%%%

\abstract{We develop the formalism of ``General gauge mediation'' for five-dimensional theories in a slice of AdS space.    A set of current correlators encodes the effect of a supersymmetry breaking hidden sector localised on the IR brane.   These current correlators provide a tree-level gaugino mass and loop-level sfermion masses on the UV brane. We also use this formalism to calculate the Casimir energy and masses for bulk hyperscalars. 

To illustrate this general construction we consider a perturbative hidden sector of generalised messengers coupled to a spurion.   
For models with large warping, we find that when the AdS warp factor $k$ is less than the characteristc mass scale $M$ of the hidden sector, the whole Kaluza-Klein tower of vector superfields propagate supersymmetry breaking effects to the UV brane. When $M$ is less than $k$,  the zero modes dominate.
}
 
%%%%%%%%%%% Introduction
\newpage

\section{Introduction}
``General gauge mediation'' (GGM) is a powerful new framework with which one can describe gauge mediation of $\mathcal{N}=1$ supersymmetry (SUSY) breaking in a model independent way \cite{Meade:2008wd}.   In GGM, the effects of breaking supersymmetry in an arbitrary and potentially strongly coupled hidden sector are captured in a small number of correlators of the gauge current supermultiplet.  These SUSY breaking effects are then mediated by the minimal supersymmetric standard model (MSSM) gauge fields to generate gaugino and scalar masses.     The GGM framework, further developed in  \cite{Buican:2008ws,deGouvea:1997tn,Intriligator:2008fr,Ooguri:2008ez,Distler:2008bt,Buican:2009vv,Argurio:2009ge,Luo:2009kf,Kobayashi:2009rn,Lee:2010kb,Intriligator:2010be,Marques:2009yu,Antoniadis:2010nj,Argurio:2010fn}, has the virtue that it clarifies which features are generic predictions of gauge mediation; for instance the existence of mass sum rules  \cite{Meade:2008wd}.  

Key to the GGM approach is a careful definition of gauge mediation, namely that  the hidden sector and the visible MSSM sector completely decouple in the limit that MSSM gauge couplings $\alpha_{i}\rightarrow 0$.   In this way one can proceed to analyse soft mass terms perturbatively in $\alpha_i$ even when the hidden sector is strongly coupled.  

This decoupling can be given a geometric context in five-dimensional models in which the hidden and visible sector are located on separate branes. The effects of supersymmetry breaking are transmitted from the hidden sector by means of MSSM gauge fields which are allowed to propagate through the bulk.     The GGM formalism has been applied to this scenario  \cite{McGarrie:2010kh} thereby generalising the Mirabelli-Peskin model \cite{Mirabelli:1997aj} to arbitrary hidden sectors.  This formalism, which we shall call GGM5d, has also recently been given a lattice (de)construction description \cite{McGarrie:2010qr,Green:2010ww,Auzzi:2010mb,Sudano:2010vt}.  

The goal of this paper is to extend the GGM5d formalism to five-dimensional theories  in a   slice of anti de Sitter space.  Such theories with warped extra dimensions are of interest for two principal reasons.     First, it is well known that warped extra dimensions can be used to generate a hierarchy of scales \cite{Randall:1999ee} and therefore such models have direct phenomenological appeal.   The second is that by means of the $AdS/CFT$ correspondence one can consider the 5D warped model as a tool with which to study certain dual strongly coupled four-dimensional CFTS (for a review see \cite{Gherghetta:2010cj,Nomura:2004zs}).   We hope that our work may have application in both these areas. 

There are, of course, a multitude of subclasses of models with warped extra dimensions \cite{Chacko:2003tf,Goldberger:2002pc,Nomura:2003qb,Nomura:2004is,Nomura:2004it,Benini:2009ff,McGuirk:2009am}.  In this paper we shall restrict our attention to those with MSSM matter located on the UV brane and a hidden sector where SUSY is broken located on the IR brane.  We do not specify the hidden sector and encode its contribution by means of current correlators. The gauge fields of MSSM propagate the effects of supersymmetry breaking between these two branes to generate soft masses as illustrated in figure \ref{warped figure}.  A difference with the four-dimensional situation is that, in general,  there is a whole  Kaluza-Klein (KK) tower of the bulk vector field contributing to the transmission of SUSY breaking.  
 
%%%%%%%%%%%%%%%%%%%%%%%%%%%%%%%%%%%%%%%%%%%%%%%%%%%%%%%%%%%%%%%%%%%%%%%%%%%%%%%%%%%%%%%%%%%%%%%%%%%%%%%%%%%%%%%%%%%%%%%%%%%%%%%%%%%%%%%%%%%%%%%%%%%%%%%%%%%%%%%%%%%%%%%%%%%%%%%%%%%%%%%%%%%%%%%%
\begin{figure}[ht]
%\vskip 5ex
\centering
\includegraphics[scale=1]{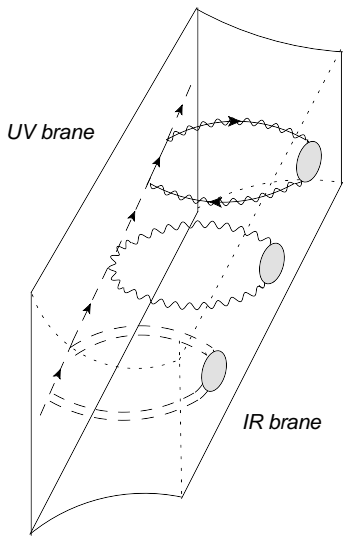}
\caption{A portrait of ``General gauge mediation'' across a warped bulk. Sfermion masses on the visible (UV) brane are generated by propagating the effects of supersymmetry breaking on the hidden (IR) brane. \label{warped figure}} 
\end{figure}
%%%%%%%%%%%%%%%%%%%%%%%%%%%%%%%%%%%%%%%%%%%%%%%%%%%%%%%%%%%%%%%%%%%%%%%%%%%%%%%%%%%%%%%%%%%%%%%%%%%%%%%%%%%%%%%%%%%%%%%%%%%%%%%%%%%%%%%%%%%%%%%%%%%%%%%%%%%%%%%%%%%%%%%%%%%%%%%%%%%%%%%%%%%%%%%%

  Within our framework,  we are able to derive general formulas for both gaugino and sfermion masses.   We are also able to accommodate bulk hypermultiplets and again give a general formula for their scalar masses.   A final application of our framework is to calculate the Casimir energy. 

To illustrate this framework we apply our general results to the case where the hidden sector has a perturbative description in terms of a SUSY breaking spurion coupled to messenger fields charged under the standard model gauge group.  We determine the structure of the current correlators in terms of these messenger fields and approximate the bulk propagators in different regimes so that we can extract approximate formulas for the gaugino and sfermion masses of the MSSM, at leading order in $\alpha_i$.  

 We explore these models with regard to four scales: the AdS warp factor $k$; orbifold length $\ell$; a SUSY breaking $F$-term and hidden sector characteristic mass scale $M$.  Assuming that  $ k\ell \gg 1$, we show that when the $F,k^{2} \ll M^2$, the full Kaluza-Klein tower of the bulk vector multiplet must be considered in the propagation of supersymmetry breaking to the UV brane, whereas in the limit that $F,M^2 \ll k^{2}$, the zero modes make the dominant contribution.

 The rest of this paper is organised as follows: In section \ref{section:Framework} we describe the field theory defined in the bulk of AdS and present the off-shell action that we require for GGM.   In section \ref{section:currents} we describe the (arbitrary)  hidden sector located on the IR brane through its current correlators.  In section  \ref{section:results} we present general results for the soft masses as well as the Casimir energy.  In section \ref{section:generalisedmess} we apply these results to the generalised messenger model. 
 %%%%%%%%%%%%%%%%%%%%%%%%%%%%%%%%%%%%%%%%%%%

\section{Off-shell warped gauge theory}\label{section:Framework}

We begin with the $AdS_{5}$ with the metric given by
\be 
\label{metric}
d s^2 = e^{-2\sigma}\eta_{\mu \nu}dx^\mu dx^\nu +dy^2,
\ee
in which  $\eta_{\mu\nu}=\textrm{diag}(-1,1,1,1)$,  $\sigma=k y$ and $1/k$ is the AdS curvature scale with mass dimension one.  From the metric, one can readily read off the  f\"unfbein to be
\be e^a_{\mu}(x,y)=e^{-\sigma}\hat{e}^a_{\mu}(x)=e^{-\sigma }\delta^a_{\mu},\quad e^5_{\mu}=e^a_{5}=0 , \quad e^{\hat{5}}_{5}=1\ . 
\ee
We are interested in describing physics on an interval of this $AdS$ space  given by  $0\le y\le \ell$ with $\ell =\pi R$.    It is helpful to think of the interval as a $\mathbb{Z}_2$ quotient of a periodic $y$ coordinate.  This construction, also known as a warped $S_1/\mathbb{Z}_2$ orbifold, is achieved by replacing $\sigma = k |y|$ in the metric (\ref{metric}) and allowing $y$ to be periodic with range $-\ell <  y\le \ell$.   The $\mathbb{Z}_2$ identification is given by $ y \sim -y$ and has fixed points at $y = 0$ and $y=\ell$ where we shall locate matter on three-branes known as the UV and IR brane respectively.

To build our GGM framework we shall need an off-shell description of the supersymmetric gauge theory living on this space. Before describing the theory in the warped orbifold lets us recapitulate the flat space case \cite{Hebecker:2001ke,Mirabelli:1997aj,McGarrie:2010kh, ArkaniHamed:2001tb}.    Since translation invariance is broken by the boundary of the interval we should anticipate only half of the 8 real supercharges in an ${\cal N}=1 $ theory in $5D$ to be preserved.   Under preserved supersymmetry  the $5D$  $\mathcal{N}=1$ vector multiplet decomposes into two multiplets which can be packaged as a chiral and a vector $4D$ $\mathcal{N}=1$  superfield given by (in Wess Zumino gauge): 
\def\th{\theta}
\def\bth{\bar{\th}}
\def\a{\alpha}
\def\s{\sigma}
\def\vth{\vartheta}
\bea
\label{multiplets}
V&=&   -\th \s^a  \bth A_a + i \bth^2 \th  \lambda - i \th^2 \bth \bar{\lambda} + \frac{1}{2} \bth^2 \th^2 D \ ,\nonumber \\
\Phi &=& \frac{1}{\sqrt{2}} \left(\Sigma +i A_5\right) + \sqrt{2} \th \chi + \th^2 F \ .
\eea
The field content of the original $5D$ vector multiplet is recovered as follows:  $A_\mu$ and $A_5$ combine to give a $5D$ gauge field $A_M$;  $\lambda$ and $\chi$ are the components of a symplectic-Majorana spinor $\Psi$; $\Sigma$ is a real scalar and $F$ and $D$ contain three real auxiliary fields in the form
\begin{equation}
D=(X^{3}-D_{5}\Sigma), \phantom{AAAA} \quad F=(X^{1}+iX^{2})\,.
\end{equation}

These  $4D$ $\mathcal{N}=1$  superfields have opposite parity under the orbifold and correspondingly obey different boundary conditions (Dirichlet for the chiral superfield and Neumann for the vector).   In terms of these superfields\footnote{Since the coupling has dimension $[g_5]= - 1/2$,   the superfields have dimension $[V] = 1/2$, $[\Phi] = 3/2$ . } the action for the gauge theory can be written as 
\be
\label{flatact}
S_{flat} = \int d^5 x  \  \frac{1}{2}\Tr \left[ \int d^2 \th  \ W^\a W_\a  + h.c. +  \frac{1}{g_5^2} \int d^4 \th\ \left(e^{-2 g_5 V }\nabla_5 e^{2g_5 V}\right)^2  \right] \ ,
\ee
where, as usual, $W_\a =- \frac{1}{4} \bar{D}^2 e^{-g_5 V} D_\a e^{g_5 V} $ and the covariant derivative acts according to 
\be
\nabla_5 ( \cdot)  = \partial_5 ( \cdot)   -  g_5 \Phi^\dag ( \cdot)   - g_5( \cdot)   \Phi \ .
\ee

To obtain the off-shell description for the warped orbifold background one may apply the procedure of ``$\th$-warping''  mentioned in \cite{Gregoire:2004nn} and developed in \cite{Bagger:2006hm,Xiong2010}.\footnote{We thank Chi Xiong and Jonathan Bagger for sharing their results with us  prior to publication.}   According to this prescription one deforms the superspace coordinates by defining $\vartheta = e^{-\frac{1}{2} \s} \th$ and supercovariant derivatives by  ${\cal D}_\a = e^{\frac{1}{2} \s} D_\a$.   Then the warped space action is given by making the replacements   $(\th, D_\a, d^5x  )\rightarrow (\vth, {\cal D}_\a ,   d^5 x \sqrt{-g})$ in the superfields  (\ref{multiplets}) and action (\ref{flatact}).  After reexpanding in the original $\th$ coordinates one finds a warped space analogue of  (\ref{flatact}) given by
\be
\label{warpact}
S_{warp} = \int d^5 x  \  \frac{1}{2}\Tr \left[ \int d^2 \th  \ W^\a W_\a  + h.c. +  \frac{e^{-2 \s}}{g_5^2} \int d^4 \th\ \left(e^{-2 g_5 V }\nabla_5 e^{2g_5 V}\right)^2  \right] \ ,
\ee
where the superfields have been warped so that they become 
\bea
\label{warpedmultiplets}
V&=&   -\th \s^a  \bth \delta_a^\mu A_\mu + i \bth^2 \th    e^{-\frac{3}{2} \s}\lambda - i \th^2 \bth e^{-\frac{3}{2} \s} \bar{\lambda} + \frac{1}{2} \bth^2 \th^2 e^{-2 \s} D \nonumber \ ,  \\
\Phi &=& \frac{1}{\sqrt{2}} \left(\Sigma +i A_5\right) + \sqrt{2} \th  e^{-\frac{1}{2} \s} \chi + \th^2 e^{-  \s} F \ .
\eea
Expanding in components one finds the kinetic terms for the vector multiplet  are given by 
\be 
 \int d^{5}x \text{Tr}\left[-\frac{1}{2}F^{\mu \nu}F_{\mu \nu} +ie^{-3\sigma}\bar{\gl}\sigma^{\mu}D_{\mu}\gl+\frac{1}{2}e^{-4\sigma}D^2 \right].
\ee  
Mass terms arise in the component expansion of the action (\ref{warpact}) from the term involving   $\partial_5 V$ when the derivative acts on the warp factor.   One finds a Dirac mass for the fermions  and upon integrating out the auxiliary $D$ field, a scalar mass  given by 
\be
\quad m_{\Psi}=\frac{1}{2}\sigma'  \, , \quad  m_{\Sigma}=-4k^2+ 2\sigma'' \, . 
 \ee
We remark that the Abelian version of this theory is related to action written in \cite{Marti:2001iw}, which makes use of a radion superfield.  
  
In what follows we shall expand these fields in terms of their eigen-modes   \cite{Gherghetta:2000qt}  which can be  summarised using the theta-warped superfields by 
\be
V = \frac{1}{\sqrt{2 \ell} }\sum_n V_n(x) f^{(2)}_n (y) \, ,  \quad \Phi = \frac{1}{\sqrt{2 \ell} } \sum_n  \Phi_n(x) g^{(4)}_n(y) \, , 
\ee
 where the even and odd modes  are given by
\bea
 f^{(s)}_{n}(y)&=&\frac{e^{s\sigma/2}}{N_{n}}\left[J_{1}\left(\frac{m_{n}e^{\sigma}}{k}\right)+b \left(m_{n}\right)Y_{1}\left(\frac{m_{n}e^{\sigma}}{k}\right)\right]\, ,  \label{ffunctions}\\ 
 g^{(s)}_{n}(y)&=&\frac{\sigma'}{k} \frac{e^{s\sigma/2}}{N_{n}}\left[J_{0}\left(\frac{m_{n}e^{\sigma}}{k}\right)+b \left(m_{n}\right)Y_{0}\left(\frac{m_{n}e^{\sigma}}{k}\right)\right]      \, , \label{gfunctions}
\eea
and obey orthonormality conditions 
 \be
\frac{1}{2\ell}\int^{\ell}_{-\ell}e^{(2-s)\sigma}f^{(s)}_{n}(y)f^{(s)}_{m}(y)dy=\delta_{nm} \label{ortho}\, ,
\ee
with similar for the odd modes. Orthornormality fixes the normalisation which in the limit $m_n\ll k$ and $k\ell \gg 1$ is given by
\be
N_m \approx \frac{e^{k\ell/2}}{\sqrt{\pi \ell m_n }} \, , 
\ee 
 and boundary conditions at both branes can be used to fix $b(m_n)= -\frac{J_0(m_n/k)}{  Y_0(m_n/k)}$ and deduce the mass spectrum by solving $b(m_n) = b(m_n e^{k \ell})$  which yields  
\be
 m_n \approx \left( n - \frac{1}{4} \right) \pi k e^{-k \ell} \, . 
\ee 
 
   \subsection{Bulk hypermultiplets}
The $\th$-warping technique can be applied to bulk hypermultiplets.  Under the orbifold action the hypermultiplet splits into two $4D$ ${\cal N}=1$ chiral superfields, $H$ of even parity and $H^c$ of odd parity  which transform under the gauge group as 
$H\to e^{-\Lambda}H$ and $H^c\to H^ce^\Lambda$.   Starting with the flat orbifold action and superfields given in \cite{Hebecker:2001ke} and following the warping procedure  one arrives at warped superfields 
\bea
H &=& H^1+\sqrt{2}e^{-\frac{1}{2}\sigma}\theta\psi_L+\theta^2 e^{-\sigma}(F_1+D_5H^2-g_{5}\Sigma H^2) \nonumber \, ,  \\
H^c &=& H^\dagger_2+\sqrt{2}\theta e^{-\frac{1}{2}\sigma} \psi_R+\theta^2 e^{-\sigma}(-F^{\dagger 2}-D_5
H^\dagger_1- g_{5} H^\dagger_1\Sigma)\, , 
\eea
and a warped action
\begin{alignat}{1}
S_{warp}^H=& \int d^5 x e^{-2\sigma}\int d^4\theta [ H^\dagger e^{2g_{5}V}H+H^c e^{-2g_{5}V}H^{c\dagger}] \nonumber \\
&\quad +\int d^5 x e^{-3\sigma} \left( \int d^2 \theta  H^c\nabla_5 H+ \int d^2\bar{\theta} H^{c\dagger}\nabla_5 H^\dagger \right)
\,.
\end{alignat}  
It should be clear from the action that the hypermultiplet will also decouple from the hidden sector in the limit $\alpha_{i}\rightarrow 0$.  In this way hypermultiplets will also follow the prescription of general gauge mediation. Starting from a massless unwarped hypermultiplet and applying $\theta$ warping, the warped hypermultiplet has a mass generated by passing a $\partial_{5}$ through an $e^{-1/2 \sigma}$ factor. This corresponds to the conformal limit $c=1/2$ for the on-shell action of \cite{Gherghetta:2000qt}.\footnote{We exclusively consider hypermultiplets at the conformal limit in this paper.} 
 The positive parity fields in $H$ have eigenfunctions given by \refe{ffunctions}, with $\alpha=1$. Similarly the negative parity fields in $H^c$ are determined by \refe{gfunctions} with $\alpha=0$. In this paper we will compute the soft mass of the zero mode scalar of $H$ which is  given by
\be
H^{(0)}(x,y)=\frac{1}{\sqrt{2\ell} }e^{\sigma}H^{(0)}(x)\; .
\ee

%%%%%%%%%%%%%%%%%%%%%%%%%%%%%%%%%%%%%%%%%%%%%%%%%%%%%%%%%%%%%%%%%%%%%%%%%%%%%%%%%%%%%%%%%%%%%%%%%%%%%%%%%%%%%%%%%%%%%%%%%%%%%%%%%%%%%%%%%%%%%%%%%%%%%%%%%%%%%%%%%%

\section{Brane localised currents} \label{section:currents}

In this section we will encode a SUSY breaking sector, localised on the IR brane at $y=\ell$, in terms of current correlators.
 
Since the vector superfield $V$ is of even parity and obeys Neumann type boundary conditions it can couple to matter charged under the gauge group localised on the boundary IR brane.   In general, we expect that the global current multiplet ${\cal J}$ should serve as a source for these interactions however we must take some care to accommodate the effects of the warping.   Starting with the flat space form of these interactions  \cite{McGarrie:2010kh} and applying the $\th$-warping technique produces boundary interactions of the form 
 \begin{equation}
S_{int}=2g_{5}\!\int\!d^5 x e^{-2\sigma} d^{4}\theta \mathcal{J} 
V \delta(y-\ell )
\end{equation}
where the warped current superfield is given by  
\bea
\mathcal{J} &=& J +    ie^{-\frac{1}{2}\sigma} \theta j  -
ie^{-\frac{1}{2}\sigma}\bar{\theta}\bar{j}  -
\theta \sigma^{a}\delta^{\mu}_{a} \bar{\theta} j_{\mu} \nonumber   \\ &  & \quad  + 
\frac{1}{2} e^{-\frac{1}{2}\sigma} \theta^2 \bar{\theta} \bar{\sigma}^{a} \delta^{\mu}_{a}
\partial_{\mu} j  - \frac{1}{2} e^{-\frac{1}{2}\s }\bar{\theta}^2 \theta 
\sigma^{a}\delta^{\mu}_{a} \partial_{\mu} \bar{j} - 
\frac{1}{4} \theta^2 \bar{\theta}^2 \square J \; ,
\eea
and $V$ is given as in (\ref{warpedmultiplets}).   In components these interaction terms read   
\begin{equation}
S_{int} = \int \!d^5 x e^{-4\sigma} g_{5}(JD- \gl j \!-
 \!\bar{\gl} \bar{j}-e^{2\sigma }j^{\mu}A_{\mu})\delta(y- \ell).
\end{equation}

The currents appearing in these expressions are not canonical in the sense they are built out of noncanonically normalised fields due to the warping of the induced metric on the IR brane.   It is helpful to instead work with canonically normalised fields and currents defined by
\be
e^{-2\sigma}J=\hat{J}\, , \quad
e^{-\frac{5}{2}\sigma}j_{\alpha}=\hat{j}_{\alpha}\, , \quad
e^{-2\sigma}j_{\mu}=\hat{j}_{\mu} \, ,   
\ee
so that the interaction terms are given by 
\begin{equation}
S_{int}=\int d^5 x  g_{5}(e^{-2\sigma} \hat{J} D- e^{-3/2\sigma}\lambda \hat{j} \!-
 \! e^{-3/2\sigma}\bar{\lambda} \hat{\bar{j}}-\hat{j}^{\mu}A_{\mu})\delta(y-\ell) \, . 
\end{equation}
With these rescalings two-point functions of canonical currents are given by  the  flat space result but with mass scales accordingly warped down.  In this way we can easily keep track of powers of the warp factor.  
 
 The contribution of the hidden sector can be found by expanding the functional integral in $g_5$  to $O(g_{5}^{2})$   following  \cite{Meade:2008wd}. Upon inserting the relation for the $D= X^3 - D_5 \Sigma$ we find that 
\begin{alignat}{1}\label{E:ChangeL}
\delta \mathcal{L}_{eff}=[&- g_{5}^{2}e^{-3k\ell}\tilde{C}_{1/2}(0) i \lambda \sigma^{\mu} \partial_{\mu} \bar{\lambda}
- g_{5}^{2}\frac {1} {4} \tilde{C}_1(0) F_{\mu\nu} F^{\mu\nu} \nonumber \\
&-g_{5}^{2}e^{-3k\ell}\frac {1}{ 2}(\hat{M} \tilde{B}_{1/2}(0) \lambda \lambda + \hat{M} \tilde{B}_{1/2}(0)\bar{\gl}\bar{\gl})\nonumber 
\\ +& e^{-4k\ell}[\frac {1}{ 2 }g_{5}^{2}\tilde{C}_0 (0)(X^{3}X^{3})+\frac {1}{ 2 }g_{5}^{2}\tilde{C}_0 (0)(D_{5}\Sigma)(D_{5}\Sigma)-g_{5}^{2}\tilde{C}_0 (0)(D_{5}\Sigma)X^{3} ] ]\ .
\end{alignat}
The $\tilde{B}$ and $\tilde{C}$ functions are related to momentum space current correlators, found below.    In position space, the current correlators of canonical currents can be expressed as
\begin{alignat}{1}
\braket{\hat{J}(x,y_{5})\hat{J}(0,y'_{5})}=&\frac{1}{x^{4}}C_0(x^{2}\hat{M}^{2})\delta(y_{5}-\ell)\delta(y'_{5}-\ell) \, , \nonumber\\
\braket{\hat{j}_\alpha(x,y_{5})\hat{\bar{j}}_{\dot\alpha}(0,y'_{5})}=&-i\sigma_{\alpha\dot\alpha}^\mu \partial_\mu(\frac{1}{x^{4}}C_{1/2}(x^{2}\hat{M}^{2}))\delta(y_{5}-\ell)\delta(y'_{5}-\ell) \, , \nonumber \\
\braket{\hat{j}_\mu(x,y_{5})\hat{j}_\nu(0,y'_{5})}=&(\partial^2\eta_{\mu\nu}-\partial_\mu \partial_\nu)(\frac{1}{x^{4}}C_1(x^{2}\hat{M}^{2}))\delta(y_{5}-\ell)\delta(y'_{5}-\ell) \, , \nonumber \\
\braket{\hat{j}_\alpha(x,y_{5})\hat{j}_\beta(0,y'_{5})}=&\epsilon_{\alpha\beta}\frac{1}{x^{5}}B_{1/2}(x^{2}\hat{M}^{2})\delta(y_{5}-\ell)\delta(y'_{5}-\ell) \label{scale} \, . 
\end{alignat}
$\hat{M}=e^{-k\ell}M$ is the characteristic mass scale of the hidden brane localised at $y=\ell$. $B_{1/2}$ is a complex function while $C_{s}$, $s=0,1/2, 1$, is real.
When supersymmetry is unbroken
\begin{equation}
C_0=C_{1/2}=C_1\;,\qquad \text{and} \qquad B_{1/2}=0\;,
\end{equation}
and since supersymmetry is restored in the UV we have
\begin{equation} \label{BCUV}
\lim_{x \rightarrow 0} C_0(x^2 \hat{M}^2)=\lim_{x \rightarrow 0} C_{1/2}(x^2 \hat{M}^2) =\lim_{x \rightarrow 0} C_1(x^2 \hat{M}^2)\;,\qquad \!\!\!\!\!\!\! \!\text{and}\!\!\!\!\!\! \!\!\qquad \lim_{x \rightarrow 0} B_{1/2}(x^2 \hat{M}^2)=0\;.
\end{equation}
 
 Defining the  Fourier transforms of $C_s$ and $B$ as 
\begin{equation}
\begin{split}
\tilde{C}_s\left(\frac{p^2}{\hat{M}^2};\frac{\hat{M}}{\Lambda}\right)&=\int d^4 x e^{ipx} \frac 1 {x^4} C_s(x^2 \hat{M}^2)\; , \\
\hat{M}\tilde{B}_{1/2}\left(\frac{p^2}{\hat{M}^2}\right)&=\int d^4 x e^{ipx} \frac 1 {x^5} B_{1/2}(x^2 \hat{M}^2)\; , 
\end{split}
\end{equation}
in which the scale $\Lambda$ is a UV cutoff regulating the integrals, allows us to express the nonzero current correlators in momentum space as\footnote{ In these expressions performing  the Fourier transforms over $y$ and $y'$ has removed the delta functions.}
\bea
\braket{\hat{J}(p,p_{5})\hat{J}(-p,p'_{5})} &=&\tilde{C}_0(p^2/\hat{M}^2)  \label{eq:c0}  \, ,\nonumber \\
\braket{\hat{j}_\alpha(p,p_{5}) \hat{\bar {j}}_{\dot\alpha}(-p,p'_{5})} &=&-\sigma_{\alpha\dot\alpha}^\mu p_\mu\tilde{C}_{1/2}(p^2/\hat{M}^2) \label{eq:c1/2}	\, , \nonumber	\\ 
\braket{\hat{j}_\mu(p,p_{5})\hat{j}_\nu(-p,p'_{5})}& =&-(p^2\eta_{\mu\nu}-p_\mu p_\nu)\tilde{C}_1(p^2/\hat{M}^2)	\label{eq:c1}\, , \nonumber	\\ 
\braket{\hat{j}_\alpha(p,p_{5})\hat{j}_\beta(-p,p'_{5})} &=&\epsilon_{\alpha\beta}\hat{M}\tilde{B}_{1/2}(p^2/\hat{M}^2)     \label{eq:b1/2} \, .	
\eea

The current correlators do not preserve incoming and outgoing $p_{5}$ momenta as the branes break Lorentz invariance in this direction.  When we come to describe the bulk fields in terms of their KK modes this has the consequence that different KK modes will get coupled to each other on the IR brane and all KK modes of the gauge field will propagate supersymmetry breaking effects. 

In what follows we shall frequently express our results in terms of the ``supertraced'' set of these current correlators 
\be
 [3\tilde{C}_1(p^2/\hat{M}^2)-4\tilde{C}_{1/2}(p^2/\hat{M}^2)+\tilde{C}_{0}(p^2/\hat{M}^2)]=\Omega \left(\frac{p^2}{\hat{M}^2} \right)\label{Omega1}\, .
\ee
The numerical coefficient in front of the $\tilde{C}_{s}$ terms in \refe{Omega1} is associated with the off-shell degrees of freedom of the bulk propagating vector multiplet and arise from taking an index contraction of the current correlators in \refe{eq:c0}. We emphasise that the effect of the induced metric on the IR brane is captured in the canonical current correlators through the warped mass scale $\hat{M}$.
%%%%%%%%%%%%%%%%%%%%%%%%%%%%%%%%%%%%%%%%%%%%%%%%%%%%%%%%%%%%%%%%%%%%%%%%%%%%%%%%%%%%%%%%%%%%%%%%%%%%%%%%%%%%%%%%%%%%%%%%%%%%%%%%%%%%%%%%%%%%%%%%%%%%%%%%%%%%%%%%%%
\section{Soft masses and vacuum energy}\label{section:results}
In this section we will give general expressions for the gaugino, sfermion and hyperscalar soft masses and the Casimir energy, in terms of the current correlators located on the IR brane.  In the next section we will explore these general expressions for a generalised messenger sector in specific limits.
%%%%%%%%%%%%%%%%%%%%%%%%%%%%%%%%%%%%%%%%%%%%%%%%%%%%%%%%%%%%%%%%%%%%%%%%%%%%%%%%%%%%%%%%%%%%%%%%%%%%%%%%%%%%%%%%%%%%%%%%%%%%%%%%%%%%%%%%%%%%%%%%%%%%%%%%%%%%%%%%%%
\subsection{Gaugino masses}
At $g^2_{5}$ order, we can extract the SUSY breaking contribution to the gaugino masses from the effective Lagrangian as 
\be
\label{gauginosoft}
\mathcal{L}_{\text{soft}}=\frac{\delta(y-\ell)}{2} g^2_{5}e^{-3k\ell}\hat{M}\tilde{B}_{1/2}(0)\lambda\lambda + \text{c.c.} 
\ee
This Majorana mass term for the gaugino  is localised on the boundary which means that  upon performing a KK decomposition this produces a term mixing all KK modes to each other given by 
\be
\mathcal{L}_{\text{soft}}  = \sum_{m, n}\frac{g^2_{4}  }{2}  \frac{1}{2} \hat{M}\tilde{B}_{1/2}(0) \lambda_n \lambda_m f_n^{(2)}(\ell)f_m^{(2)}(\ell) +  \text{c.c.}   
\ee
in which we have expressed the answer in terms of the dimensionless $4D$ gauge coupling  $g^2_{5}= g^2_{4} \ell$.

  In addition to these masses there are also Dirac type Kaluza-Klein masses which mix the positive parity gaugino modes with the negative parity bulk fermion modes.  In general, one must therefore take into account both to understand the gaugino spectrum.     In practice the easiest way to find the gaugino masses is to include the contribution from   \refe{gauginosoft} in the boundary conditions placed on the KK mode expansions as shown in the appendix of  \cite{Marti:2001iw}, similar to the flat case in \cite{ArkaniHamed:2001mi}.   
  
  %%%%%%%%%%%%%%%%%%%%%%%%%%%%%%%%%%%%%%%%%%%%%%%%%%%%%%%%%%%%%%%%%%%%%%%%%%%%%%%%%%%%%%%%%%%%%%%%%%%%%%%%%%%%%%%%%%%%%%%%%%%%%%%%%%%%%%%%%%%%%%%%%%%%%%%%%%%%%%%%%%
\subsection{Sfermion masses}
The sfermion masses of the MSSM can be determined from the $\tilde{C}_{s}$ current correlators of the SUSY breaking fields on the IR brane and the propagation of this breaking by the vector multiplets in the bulk, up to the UV brane as shown in figure \ref{figure2}.    The full momentum dependence of the current correlators should be taken into account as they form part of a loop on the scalar propagator.   The vertex couplings can all be obtained from expanding out a canonical K\"ahler potential for a chiral superfield.    To massage the answers into their final form  one makes use of the  representations   \cite{McGarrie:2010kh,Mirabelli:1997aj}
\be
\delta (0) =\frac{1}{2\ell} \sum_{n} 1   \, , \quad  0 = \sum_n (-1)^n \, .  \label{delta0}
\ee
Judicial use of the later identity allows us to replace factors of $m_n^2$ which occur in the rightmost diagram of  figure \ref{figure2} with $p^2$. 
%%%%%%%%%%%%%%%%%%%%%%%%%%%%%%%%%%%%%%%%%%%%%%%%%%%%%%%%%%%%%%%%%%%%%%%%%%%%%%%%%%%%%%%%%%%%%%%%%%%%%%%%%%%%%%%%%%%%%%%%%%%%%%%%
\begin{figure}[ht]
%\vskip 5ex
\centering
\includegraphics[scale=0.8]{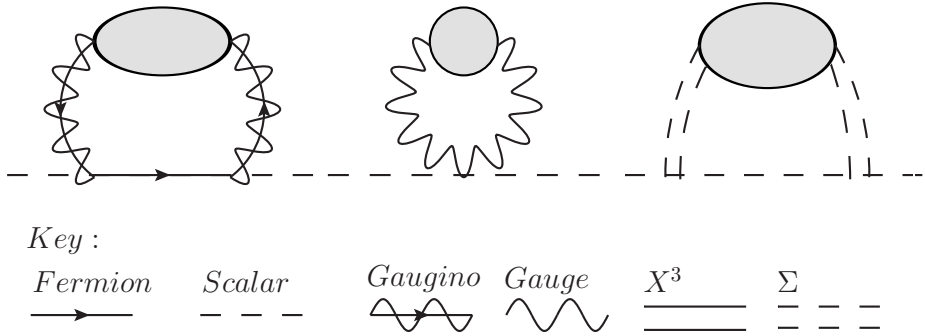}
\caption{The graphical description of the contributions of the two-point functions to the soft sfermion masses. The ``blobs'' represent current correlators localised on the IR brane at $y=\ell$.  The scalar external legs are the sfermions located on the UV brane.  The first diagram represents the current correlator $\braket{j_{\alpha}\bar{j}_{\bar{\alpha}}}$ being mediated by the bulk gaugino $\gl$ from the IR brane to the UV brane at $y=0$.  The second diagram represents mediation of $\braket{j^{\mu} j^{\nu}}$ due to the bulk gauge boson and the final diagram represents mediation of the scalar current correlator $\braket{JJ}$ due to the negative parity bulk scalar $\Sigma$. This is the complete supertraced combination of diagrams for gauge mediation \cite{Meade:2008wd,McGarrie:2010kh}.}
\label{figure2}
\end{figure}
%%%%%%%%%%%%%%%%%%%%%%%%%%%%%%%%%%%%%%%%%%%%%%%%%%%%%%%%%%%%%%%%%%%%%%%%%%%%%%%%%%%%%%%%%%%%%%%%%%%%%%%%%%%%%%%%%%%%%%%%%%%%%%%%
 
 The final result may be written by first defining a positive parity bulk field in the AdS background, propagating from $y=y$ to $y=y'$ \cite{McGarrie:2010kh,Ichinose:2006en,Mirabelli:1997aj,Ichinose:2007zz}
\be
\tilde{G}(y,y')=\frac{1}{2\ell}\sum_{n} \frac{f^{(2)}_{n}(y)f^{(2)}_{n}(y')}{p^2+m^2_{n}} \label{gprop}\, .
\ee
 Then the sfermion mass formula is found to be 
\begin{equation}
m_{\tilde{f}}^2= \sum _r g_{r(5d)}^4  c_2(f;r)E_r
\end{equation}
where
\begin{equation}
E_r= -\! \!\int\! \frac{d^4p}{ (2\pi)^4}\tilde{G}(0,\ell)\tilde{G}(0,\ell) p^2 \Omega^{(r)} \left(\frac{p^2}{\hat{M}^2} \right). \label{P1}
\end{equation}
$r=1,2,3,$ refer to the gauge groups $U(1),SU(2), SU(3)$. $c_2(f;r)$ is the quadratic Casimir for the representation of sfermion $\tilde{f}$, the scalar  in question. 

  As was discussed in \cite{McGarrie:2010kh}, it is interesting to consider the contributions to these masses in different regimes of warping $k$ relative to $M^2, F$.   Since we have that $m_{KK} \approx k^2 e^{-2kl} $ we can see that in the small or intermediate regimes  $k^{2} \ll F,M^2$ or $ F \le k^{2}\ll M^2$ the Kaluza-Klein modes contribute to the mediation of supersymmetry breaking effects across the bulk whereas for large warping, $F ,M^2 \ll k^{2}$, only the zero modes will contribute significantly and we will have an effective $4D$ model with AdS effects.

If  only the zero modes contribute to propagation we may truncate the KK tower. One may then write the zero mode eigenfunctions in terms of gauge boson zero mode eigenfunctions.  The zero mode gauge boson eigenfunctions are flat and one obtains
\begin{equation}
E_r= -\! \!\int\! \frac{d^4p}{ (2\pi)^4}\frac {1}{\ell^{2}} \!  \frac{1}{p^2} \Omega^{(r)}\left(\frac{p^2}{\hat{M}^2} \right). \label{P12}
\end{equation}
Defining $g^2_{(5d)}/\ell=g^2_{(4d)}$, we recover exactly the four-dimensional GGM answer \cite{Meade:2008wd}, with $F$ and $M$ replaced with $\hat{F}$ and $\hat{M}$. This is an effective four-dimensional limit, however as the current correlators are a function of $\hat{M}$, we will still find some suppression due to the warp factor.
%%%%%%%%%%%%%%%%%%%%%%%%%%%%%%%%%%%%%%%%%%%%%%%%%%%%%%%%%%%%%%%%%%%%%%%%%%%%%%%%%%%%%%%%%%%%%%%%%%%%%%%%%%%%%%%%%%%%%%%%%%%%%%%%%%%%%%%%%%%%%%%%%%%%%%%%%%%%%%%%%%
\subsection{Hypermultiplet scalar masses}
\begin{figure}[ht]
%\vskip 5ex
\centering
\includegraphics[scale=0.8]{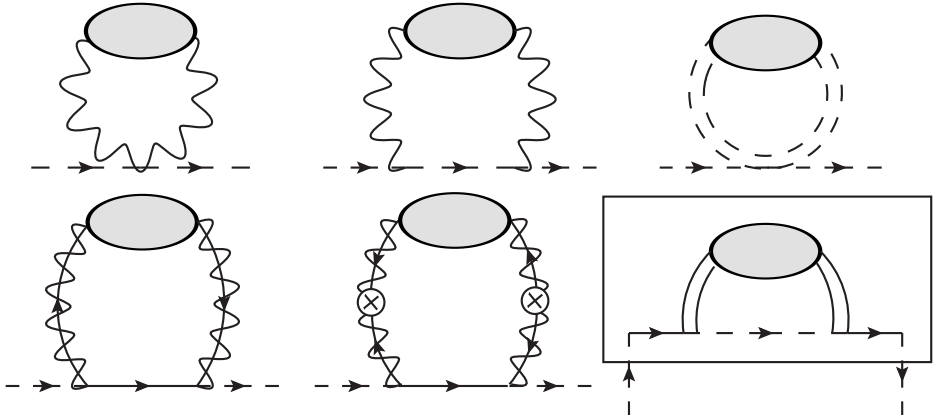}
\caption{The graphical description of the contributions of the two point functions to the hypermultiplet scalar masses. The current correlators (blobs) are located on the IR brane at $y=\ell$.  The external hyperscalar zero mode legs are in the bulk and one must integrate over all possible positions of the external legs.} \label{figure3}
\end{figure}
%%%%%%%%%%%%%%%%%%%%%%%%%%%%%%%%%%%%%%%%%%%%%%%%%%%%%%%%%%%%%%%%%%%%%%%%%%%%%%%%%%%%%%%%%%%%%%%%%%%%%%%%%%%%%%%%%%%%%%%%%%%%%%%%%%%%%%%%%%%%%%%%%%%%%%%%%%%%%%%%%%
The supersymmetry breaking masses of a bulk hypermultiplet scalar zero mode may be computed using the general gauge mediation prescription as it decouples from the hidden sector in the limit $\alpha_{i}\rightarrow 0$. The diagrams needed to compute the soft mass are found in figure \ref{figure3}.  When computing the mass, the second diagram vanishes due to transversality. To compute the diagrams, one again takes the current correlators to be brane localised and do not preserve the incoming and outgoing $m_{n}$ eigenmass. The external hypermultiplet legs with mass $m_{n}=0$ eigenfunctions must be specified when computing the diagrams and the vertices to the external hyperscalar must be integrated over all of $y$.  The orthornormality condition \refe{ortho} is used in this integration over $y$.  The right column of diagrams may be collected together by use of \refe{delta0}.  The rectangle on the last diagram signifies that this diagram is completely localised on the IR brane.  The diagram with the symbol $\otimes$, represents a bulk propagator that couples the positive parity fermion $\gl$, to the negative parity fermion $\chi$. 

We find the zero mode hyperscalar soft mass is given by
\begin{equation}
E_r= -\! \!\int\! \frac{d^4p}{ (2\pi)^4}\frac {1}{2\ell^2}\sum_{n} \!  \frac{f^{(2)}_n (\ell)f^{(2)}_n(\ell)}{p^2+m_{n}^{2}}\frac{p^{2}}{p^2+m_{n}^{2}} \Omega^{(r)}\left(\frac{p^2}{\hat{M}^2} \right). \label{P11}
\end{equation}
%%%%%%%%%%%%%%%%%%%%%%%%%%%%%%%%%%%%%%%%%%%%%%%%%%%%%%%%%%%%%%%%%%%%%%%%%%%%%%%%%%%%%%%%%%%%%%%%%%%%%%%%%%%%%%%%%%%%%%%%%%%%%%%%%%%%%%%%%%%%%%%%%%%%%%%%%%%%%%%%%%
It should be noted that this equation has only a single sum as the hyperscalar vertices do preserve incoming and outgoing $m_{n}$, unlike for the branes which do not.
The momentum integral is UV divergent as expected \cite{Puchwein:2003jq,Georgi:2000ks}: as we must integrate over all of $y$ for the sewing points of externals hyperscalar legs, there is no finite separation between the bulk hyperscalar and the IR brane to suppress large momentum contributions in the loop.
%%%%%%%%%%%%%%%%%%%%%%%%%%%%%%%%%%%%%%%%%%%%%%%%%%%%%%%%%%%%%%%%%%%%%%%%%%%%%%%%%%%%%%%%%%%%%%%%%%%%%%%%%%%%%%%%%%%%%%%%%%%%%%%%%%%%%%%%%%%%%%%%%%%%%%%%%%%%%%%%%%
\subsection{Vacuum energy}
The propagation of supersymmetry in the bulk also produces a nonzero vacuum energy. Each vacuum diagram is generated by forming a closed loop from one end of the current correlator to the other, with the fields that mediate the effects of those current correlators.  There are in fact four diagrams: a closed loop with the gauge boson, gaugino, $\Sigma$ and $X^3$.  The field $X^3$ is non propagating and never leaves the IR brane. The diagram of $X^3$ combines with that of the scalar $\Sigma$ to generate the contribution from the current correlator $C_{0}$. The vacuum energy density is 
\begin{equation}
\mathcal{E}=\sum_{r} \frac{1}{2} g^{2}_{5}  d_{G_{r}} \int \frac{d^{4}p}{(2\pi)^{4}}\tilde{G}(\ell,\ell) p^2\Omega^{(r)}\left(\frac{p^2}{\hat{M}^2} \right). \label{Vacuum1}
\end{equation}
$d_{G}$ is the dimension of the adjoint representation of the gauge group labeled by $r$. The vacuum energy is UV divergent. In the next section we will demonstrate how the Casimir energy may be extracted from this formula, in the $k\ell\gg 1$ limit. 
%%%%%%%%%%%%%%%%%%%%%%%%%%%%%%%%%%%%%%%%%%%%%%%%%%%%%%%%%%%%%%%%%%%%%%%%%%%%%%%%%%%%%%%%%%%%%%%%%%%%%%%%%%%%%%%%%%%%%%%%%%%%%%%%%%%%%%%%%%%%%%%%%%%%%%%%%%%%%%%%%%
\section{Generalised messenger sector}\label{section:generalisedmess}
In this section we give a concrete description of matter content of the IR SUSY breaking brane following the construction of \cite{Martin:1996zb}.  We consider two sets of $SU(N)$ vectorlike Chiral superfield messengers $\phi_{i},\tilde{\phi}_{i}$ coupled to a SUSY breaking spurion $X= M + \theta^2 e^{-\sigma} F $. Generalisations to arbitrary hidden sectors are just a straightforward application of the results of  \cite{McGarrie:2010kh,Marques:2009yu}.  The superpotential, which is localised on the IR brane, is 
\be W  =X \eta_{i}\
\phi_i \ti \phi_i . \label{superpotential2}\ee
In principle $\eta_{ij}$ is a generic matrix which may be diagonalised to its eigenvalues $\eta_{i}$ \cite{Martin:1996zb}.  The index $i$ labels the number of messengers from $1$ to $N$.  The messengers on the SUSY breaking brane will couple to the bulk vector superfield as
\be \delta {\cal L} = \int d^2\theta d^2\bar\theta e^{-2\sigma}\left(\phi^\dag_i
e^{2 g_{5} V^a T^a} \phi_i + \ti\phi^\dag_i e^{-2 g_{5} V^a T^a}
\ti\phi_i\right) + \left(\int d^2\theta\ e^{-3\sigma}  W +
c.c.\right) .\label{hiddensector} \ee
We can extract the multiplet of currents from the kinetic terms in the above Lagrangian.  It is easiest to work in canonical fields and canonical currents as discussed in section \ref{section:currents}.  It is also useful to absorb warp factors into the mass scales such that
\be
e^{-k\ell}M=\hat{M}\, , \quad
e^{-2k\ell}F=\hat{F}\, , \quad
e^{-k\ell}k=k^* \, .   
\ee

%%%%%%%%%%%%%%%%%%%%%%%%%%%%%%%%%%%%%%%%%%%%%%%%%%%%%%%%%%%%%%%%%%%%%%%%%%%%%%%%%%%%%%%%%%%%%%%%%%%%%%%%%%%%%%%%%%%%%%%%%%%%%%%%%%%%%%%%%%%%%%%%%%%%%%%%%%%%%%%%%%%%%
\subsection{Gaugino masses}
The gaugino soft mass matrix is given by 
\be
\mathcal{L}_{\text{soft}}  = \sum_{m, n}\frac{g^2_{4}  }{2} \frac{1}{2}  \hat{M}\tilde{B}_{1/2}(0) \lambda_n \lambda_m f_n^{(2)}(\ell)f_m^{(2)}(\ell) + \text{c.c.}   
\ee
It should be clear that there is a mass term for coupling all gaugino modes of the Kaluza-Klein tower to all other modes.  Consider just the zero mode portion of this matrix given by 
 \be
 \frac{1}{2} M_{\lambda_0}  \lambda_0 \lambda_0 \equiv  \frac{g^2_{4}}{2}  \frac{1}{2}   \hat{M}\tilde{B}_{1/2}(0)  \lambda_0 \lambda_0 f_0^{(2)}(\ell)f_0^{(2)}(\ell)  =
 \frac{g^2_{4}}{2}   \frac{1}{2}  \hat{M}\tilde{B}_{1/2}(0)  \lambda_0 \lambda_0  
  \ee
 where the   equality is due to the fact that the zero mode for the gauge field is flat, $f^{(2)}_0(y) = 1$.  
      This allows us to establish for the minimal messenger model the scale associated with the gaugino mass 
 \be 
  M^{(r)}_{\lambda_0}  =  \frac{\alpha_r}{4\pi} \Lambda_G\, , \quad \Lambda_{G}=\sum_{i=1}^{N}[e^{-k\ell}\frac{d_{r}(i) F}{M }2 g(x_{i})], 
\ee
 in which have used the result for $ \hat{M}\tilde{B}_{1/2}(0)$ obtained in flat space analogue  \cite{McGarrie:2010kh}  but with the mass  warped accordingly.   The label $r=1,2,3$ refers to the gauge groups $U(1),SU(2),SU(3)$, $d_{a}(i)$ is the Dynkin index of the representation $i$ and
\be 
g(x)= \frac{(1-x)\log(1-x)+(1+x)\log(1+x)}{ x^2}
 \ee
where  $x_{i}=\frac{\hat{F}}{\eta_{i}\hat{M}^2}=\frac{F}{\eta_{i}M^2}$ and $g(x)\sim 1$ for small $x$ \cite{Martin:1996zb}.  

If we wish to consider the full mass matrix we can approximate $f^{(2)}_{n >0}(\ell) = \sqrt{2 k\ell } (-1)^n$  giving rise to 
\be
\frac{1}{2} M_{\lambda_{n,m}}  \lambda_n\lambda_m  =  \frac{g^2_{4}}{2}  (-1)^{(n+m)}  k\ell  \hat{M}\tilde{B}_{1/2}(0)  \lambda_n \lambda_m  \, .
\ee
From this one can see the soft mass contribution associated with the nonzero modes is a factor    $k\ell $ greater than those of the zero mode alone.    
%%%%%%%%%%%%%%%%%%%%%%%%%%%%%%%%%%%%%%%%%%%%%%%%%%%%%%%%%%%%%%%%%%%%%%%%%%%%%%%%%%%%%%%%%%%%%%%%%%%%%%%%%%%%%%%%%%%%%%%%%%%%%%%%%%%%%%%%%%%%%%%%%%%%%%%%%%%%%%%%%% 
\subsection{Sfermion masses}
The sfermion masses are sensitive to the warping $k^*=k e^{-k\ell}$. Due to the complicated nature of the bulk propagators, we will only comment on the limit when $k\ell$ is large. For the KK modes to contribute, we require that $k$ is small such that $k^2 \ll F,M^2$ or intermediate such that $F \le k^{2}  \ll 
M^2$. We take the exact results of \refe{P1} with \refe{gprop}.  As demonstrated in Appendix \ref{appendixb},  we approximate the full KK tower of propagators with the eigenfunctions and eigenstates of the heavier modes, which are less localised to the IR brane so will contribute most to the propagation across the bulk.  We then complete a Matsubara summation of the tower of modes. We find the propagator to be
\be
\tilde{G}(0,\ell)\sim \frac{ 2e^{kl/2}(-1)^n}{p \sinh(p/k^*)} ,
\ee
which gives
\be 
E_r \simeq -\! \!\int\! \frac{d^4p}{ (2\pi)^4} \frac{4 e^{k\ell}}{\sinh^2(p/k^*)}  \Omega^{(r)}\left(\frac{p^2}{\hat{M}^2} \right).
\ee
Physically, the tower of KK modes suppresses large momenta contributions from the two point functions on the IR brane, which can be seen from the behaviour of $\sinh^2(p/k^*)$.  In this regime we may expand the current correlators for small momenta as found in \cite{McGarrie:2010kh}, valid when $\frac{p^2}{M^2}\rightarrow 0$:
\be
 \Omega^{(r)}\left(\frac{p^2}{\hat{M}^2} \right) \approx -\frac{1}{(4\pi)^2}\frac{2 d_{r}}{3}x^2 h(x) + O(p^2)
\ee  
with
\be h(x) =\frac{3}{2}[\frac{4+x-2x^2}{x^4}\log(1+x)+\frac{1}{x^2}] + (x\rightarrow -x).
\ee
$h(x)$ for $x<0.8$ can be reasonably approximated by $h(x)=1$ \cite{McGarrie:2010kh}.  In this limit, the function is independent of $p^2$.  Finally, the momentum integral is evaluated and the sfermions masses can then be written as 
\be
 m_{\tilde{f}}^2 \sim 2 C_{\tilde{f}}^r\sum^{3}_r (\frac{\alpha_{r}}{4\pi})^2 \sum_{i}^{N} 4(k^2 \ell)^2 e^{-3k\ell}\zeta(3) d_{r}(i)|\frac{F}{\eta_{i}M^2}|^2 h(x_{i}).
\ee
$C_{\tilde{f}}^r$ is the quadratic Casimir of the $\ti f$ scalar in question, in the gauge group $r$. The sfermion scale $\Lambda_S^2$ is
\be 
\Lambda_S^2 \sim  2 \sum_{i=1}^{N}\left(\frac{\ell k^2 }{|\eta_{i}M |}\right)^2  \left|\frac{F}{M}\right|^2  e^{- 3k\ell}     \zeta(3)
d_{r}(i) h(x_i)\label{LSMGM}.\ee
Next, we may turn to the limit $F,M^2 \ll k^2$.  In this limit only the zero modes contribute significantly to the mediation across the bulk.  We again start from \refe{P1} and keep only the zero modes in the bulk propagators.  One can rewrite this in terms of the zero mode gauge boson eigenfunctions, which are flat \cite{Pomarol:1998sd,Gherghetta:2000kr,Marti:2001iw}.  We arrive at
\begin{equation}
E_r= -\! \!\int\! \frac{d^4p}{ (2\pi)^4}\frac {1}{\ell^{2}}\frac{1}{p^{2}} [3\tilde{C}_1^{(r)}(p^2/\hat{M}^2)-4\tilde{C}_{1/2}^{(r)}(p^2/\hat{M}^2)+\tilde{C}_{0}^{(r)}(p^2/\hat{M}^2)].
\end{equation}
Similarly, one may use the full answer \refe{P1} and expand the current correlators in the limit $\frac{p^2}{M^2}\rightarrow \infty$ as was explored in the appendix of \cite{Mirabelli:1997aj}. The above equation has been evaluated before \cite{Martin:1996zb,Meade:2008wd,Marques:2009yu}, which we rescale by use of $\hat{F}$ and $\hat{M}$. The result is 
\be
 m_{\tilde{f}}^2 = 2 \sum^{3}_r C_{\tilde{f}}^r (\frac{\alpha_{r}}{4\pi})^2 \sum_{i}^{N} e^{-2k\ell} d_{r}(i)|\frac{F}{M}|^2f(x_{i}) .\label{4dlimit}
\ee
$f(x_{i})$ is given by 
\be
f(x)=\frac{1+x}{x^2}[\log (1+x)-2\text{Li}_{2}(\frac{x}{[1+x]})+\frac{1}{2}\text{Li}_{2}(\frac{2x}{[1+x]})]+(x\rightarrow -x) .
\ee
For $x<1$, $f(x)\sim 1$. The additional factor of $e^{-2k\ell}$ arises from the dimensionful ratio of $|\frac{\hat{F}}{\hat{M}}|^2$, in the supertrace of $\tilde{C}$ terms. The factors of warping cancel in the ratio $\Lambda^2_{G}/\Lambda^2_{S}$ in this ``$4D$ limit''.  These results show that depending on the ratio of $k$ to $M$, one may obtain an ordinary ``gauge mediated'' result or ``gaugino mediated'' spectrum.  
%%%%%%%%%%%%%%%%%%%%%%%%%%%%%%%%%%%%%%%%%%%%%%%%%%%%%%%%%%%%%%%%%%%%%%%%%%%%%%%%%%%%%%%%%%%%%%%%%%%%%%%%%%%%%%%%%%%%%%%%%%%%%%%%%%%%%%%%%%%%%%%%%%%%%%%%%%%%%%%%%%%%%%%%%%%%%%%%%%%%%%%%%%%%%%%%%%
%%%%%%%%%%%%%%%%%%%%%%%%%%%%%%%%%%%%%%%%%%%%%%%%%%%%%%%%%%%%%%%%%%%%%%%%%%%%%%%%%%%%%%%%%%%%%%%%%%%%%%%%%%%%%%%%%%%%%%%%%%%%%%%%%%%%%%%%%%%%%%%%%%%%%%%%%%%%%%%%%%%%%
\subsection{Hyperscalar mass}
We now focus on computing the zero mode hyperscalar soft mass. We start with the exact result found in \refe{P11} and approximate the warped propagator KK mode eigenfunctions by 
\be 
f^{(2)}_{n}(\ell)f^{(2)}_{n}(\ell)\simeq 4 (k \ell) ,
\ee
to find 
\begin{equation}
m_{H_{0}}^2= \sum _r g_{r(5d)}^4 c_2(f;r)D_r
\end{equation}
where, after performing a Matsubara frequency summation,
\be 
D_r \sim \! -\!\int\! \frac{d^4p}{ (2\pi)^4}\frac{2(k\ell)}{\ell^2 k^*}\frac{\coth (p/k^*)+ p/k^*\text{csch}^2(p/k^*)  }{2 p}  \Omega^{(r)}\left(\frac{p^2}{\hat{M}^2} \right). \label{good}
\ee
$D_{r}$ is UV divergent due to the loop in the hypermultiplet diagrams not always being spatially separated by the interval. We would like to extract the $\slashed{D}_{r}$, the $k^*$ dependent part of $D_{r}$. Upon subtracting the UV limit of the integrand, we find
\be
 \slashed{D}_r= -\! \!\int\! \frac{d^4p}{ (2\pi)^4}\frac {k}{\ell k^* }[\frac{\coth(p/k^*)+(p/k^*)\text{csch}^2(p/k^*)-1 }{p}] \Omega^{(r)}\left(\frac{p^2}{\hat{M}^2} \right). 
\ee
We use the expansion of current correlators in the limit $\frac{p^2}{M^2}\rightarrow 0$ found above, to obtain 
\be
 m_{\tilde{H}}^2 \sim \frac{2}{3} C_{\tilde{f}}^r\sum^{3}_r (\frac{\alpha_{r}}{4\pi})^2 \sum_{i}^{N} 4k^2  (k\ell) e^{-2k\ell}\zeta(3) d_{r}(i)|\frac{F}{\eta_{i}M^2}|^2 h(x_{i}).
\ee
This result is less warped than the sfermion soft mass \refe{LSMGM}.  Now we comment on the limit of $M\ll k$.  In this AdS $4D$ limit one obtains the same result as for the sfermion masses \refe{4dlimit}.
%%%%%%%%%%%%%%%%%%%%%%%%%%%%%%%%%%%%%%%%%%%%%%%%%%%%%%%%%%%%%%%%%%%%%%%%%%%%%%%%%%%%%%%%%%%%%%%%%%%%%%%%%%%%%%%%%%%%%%%%%%%%%%%%%%%%%%%%%%%%%%%%%%%%%%%%%%%%%%%%%%
%%%%%%%%%%%%%%%%%%%%%%%%%%%%%%%%%%%%%%%%%%%%%%%%%%%%%%%%%%%%%%%%%%%%%%%%%%%%%%%%%%%%%%%%%%%%%%%%%%%%%%%%%%%%%%%%%%%%%%%%%%%%%%%%%%%%%%%%%%%%%%%%%%%%%%%%%%%%%%%%%%%%%
\subsection{Vacuum energy}
The vacuum energy can be computed starting from \refe{Vacuum1}. The vacuum energy is UV divergent. To obtain the finite part, one may extract the UV limit of the momentum integrand as was carried out for the hyperscalar masses above, or one may use a contour trick in the matsubara summation, outlined in \cite{Mirabelli:1997aj}. Either way leads to the same answer.  For definiteness we use the contour trick and obtain the Casimir energy in the limit of $k\ll M$,
\begin{equation}
\mathcal{E}_{\text{Casimir}}\sim \sum_{r}g^{2}_{5} d_{G_{r}}\int \frac{d^{4}p}{(2\pi)^{4}}\frac{k \ell e^{2k \ell} p}{ e^{2p /k^*}-1} \Omega^{(r)}\left(\frac{p^2}{\hat{M}^2} \right).\label{casimirenergy}
\end{equation}
Evaluating this for the case of minimal messengers, we find
\be
\mathcal{E}_{\text{Casimir}}\sim  - \frac{1}{2} \sum_{r}\sum_{i}^{N}\frac{g^{2}_{4d (r)}  d_{G_{r}} d_{r}(i) \zeta (5)}{128 \pi^4} (k\ell)k^4 e^{-4k \ell}\left|\frac{F}{\eta_{i}M^2}\right|^2 h(x_{i}) \label{casimirenergy2}. 
\ee
The warped factor $e^{-4k \ell}$ dominate this result as $k\ell\gg 1$ and we find effectively zero Casimir energy in this regime.  In the $4D$ AdS limit we find
\be
\mathcal{E}_{\text{Casimir}} =-\frac{1}{2} \sum_{i}\sum_{r}\frac{4 g^2_{4d(r)} } {(4\pi)^4} d_{g_{r}} d_{r}(i)|\eta_{i} F|^2 e^{-4k\ell} \log (\frac{k}{\eta_{i}M}) 
\ee
which is just as suppressed by the warp factor $e^{-4k \ell}$.  This negative contribution to the Casimir energy is solely from $F$ term breaking of supersymmetry.  It would be interesting to explore the contributions from $D$ term breaking as they have positively signed contributions to the Casimir energy \cite{Mirabelli:1997aj}.  Further, one may include supergravity contributions and explore minimising the total vacuum energy.
%%%%%%%%%%%%%%%%%%%%%%%%%%%%%%%%%%%%%%%%%%%%%%%%%%%%%%%%%%%%%%%%%%%%%%%%%%%%%%%%%%%%%%%%%%%%%%%%%%%%%%%%%%%%%%%%%%%%%%%%%%%%%%%%%%%%%%%%%%%%%%%%%%%%%%%%%%%%%%%%%%
\section{Discussion and conclusion}
General gauge mediation is a powerful model independent framework for gauge mediated supersymmetry breaking.  In this work we have shown how the GGM approach can be applied to five-dimensional warped models.  We started with an off-shell $\mathcal{N}=1$ $5D$ bulk super Yang-Mills action in a slice of $AdS_{5}$ with $k\ell\gg 1$.  We located a supersymmetry breaking hidden sector on the IR brane, located chiral superfields on the UV  brane and hypermultiplets in the bulk.  Using the construction of general gauge mediation we encoded generic supersymmetry breaking effects in terms of current correlators.  We used the current correlators to compute bulk gaugino masses, scalar masses of the UV localised chiral superfields,   bulk hyperscalar masses and finally computed the Casimir energy.  We then specified the hidden sector to be a general messenger sector coupled to a spurion. This allowed us to perturbatively evaluate the current correlators. Using some simplifying assumptions, to sum up the effects of the Kaluza-Klein tower of bulk vector superfields, we propagated the effect of these current correlators and gave approximate formulas for the SUSY breaking masses and vacuum energy.  

There are several interesting features of this model. We find a natural geometric interpretation of the soft terms being hierarchically small as all soft mass scales arise with at least one factor of $e^{-k\ell}$.  In the $4D$ limit of our model, when $M\ll k$, the masses are warped down but the warp factors cancel in the ratio $\Lambda^2_{G}/\Lambda^2_{S}$ and thus the model behaves like a typical gauge mediated scenario.   However in the other limit $k\ll M$ the model demonstrates a quite different behaviour.   Although we expect $\Lambda_G$ to remain of the same order, $\Lambda_S^2$ acquires a suppression by both a dimensionless ratio, as happens in the non-warped case, {\it and} an additional warp factor $e^{-k\ell}$.  In this way it is feasible to obtain not only a hierarchy  between SUSY breaking and Planck scales but also between different soft mass scales.

Two recent models which fit into the framework of supersymmetry breaking on an IR brane by a hidden sector are  \cite{Dudas:2007hq} and \cite{Abel:2010uw}. It would be interesting to see the soft mass formulas of this paper applied to these models. The phenomenology of pure general gauge mediation in four-dimensions has been explored in \cite{Abel:2009ve,Abel:2010vb} and a similar investigation is warranted for five-dimensional models.  It would be useful to compare numerical calculations, especially of the approximations of the Matsubara sum of the full Kaluza-Klein tower, with the approximate results of this paper. We would also like to understand if it is possible to interpolate between the two limits, explored in this paper, using the full Kaluza-Klein tower, as has been possible in a simpler lattice type model in which the effective KK tower has only two states \cite{Auzzi:2010mb}.
%%%%%%%%%%%%%%%%%%%%%%%%%%%%%%%%%%%%%%%%%%%%%%%%%%%%%%%%%%%%%%%%%%%%%%%%%%%%%%%%%%%%%%%%%%%%%%%%%%%%%%%%%%%%%%%%%%%%%%%%%%%%%%%%%%%%%%%%%%%%%%%%%%%%%%%%%%%%%%%%%%%%%%%%%%
\paragraph{Acknowledgments} 
We would like to thank Steven Thomas, Rodolfo Russo, Daniel Koschade, Alberto Mariotti and Chi Xiong for useful discussions.  MM is funded by STFC.  During this research DCT was supported in part by STFC and by the Belgian Federal Science Policy Office through the Interuniversity Attraction Pole IAP VI/11 and by FWO-Vlaanderen through project G011410N.
%%%%%%%%%%%%%%%%%%%%%%%%%%%%%%%%%%%%%%%%%%%%%%%%%%%%%%%%%%%%%%%%%%%%%%%%%%%%%%%%%%%%%%%%%%%%%%%%%%%%%%%%%%%%%%%%%%%%%%%%%%%%%%%%%%%%%%%%%%%%%%%%%%%%%%%%%%%%%%%%%%%%%%%%%%
\appendix

%%%%%%%%%%%%%%%%%%%%%%%%%%%%%%%%%%%%%%%%%%%%%%%%%%%%%%%%%%%%%%%%%%%%%%%%%%%%%%%%%%%%%%%%%%%%%%%%%%%%%%%%%%%%%%%%%%%%%%%%%%%%%%%%%%%%%%%%%%%%%%%%%%%%%%%%%%%%%%%%%% %%%%%%%%%%%%%%%%%%%%%%%%%%%%%%%%%%%%%%%%%%%%%%%%%%%%%%%%%%%%%%%%%%%%%%%%%%%%%%%%%%%%%%%%%%%%%%%%%%%%%%%%%%%%%%%%%%%%%%%%%%%%%%%%%%%%%%%%%%%%%%%%%%%%%%%%%%%%%%%%%%
 \section{Limits on the bulk propagator} \label{appendixb}
All Kaluza-Klein modes may propagate supersymmetry breaking across the interval.  To discern which modes contribute most, it is important to determine the limiting behaviour of the bulk eigenfunctions, as they change with mass eigenstates $m_{n}$   \cite{Ichinose:2006en}.
First we focus on the positive parity eigenfunctions 
\be 
f^{(2)}_{n}(y)=\frac{e^{\sigma}}{N_{n}}\left[J_{1}\left(\frac{m_{n}e^{\sigma}}{k}\right)+b(m_{n})Y_{1}\left(\frac{m_{n}e^{\sigma}}{k}\right)\right].
\ee
Let us first evaluate the function at $y=\ell$ in the mass regime $m_{n}\gg k$ and take  $x=m_{n}e^{\sigma}/k$.  In this regime $m_{n}\simeq n\pi k^*$ with very large $n$. Using the identities valid for large $x$,
\be
J_{1}(x)\simeq\left(\frac{2}{\pi x}\right)^{1/2}\cos (x-3\pi/4)=  \left(\frac{2}{\pi x}\right)^{1/2}\sin (x-\pi/4)
\ee
\be
Y_{1}(x))\simeq\left(\frac{2}{\pi x}\right)^{1/2}\sin (x-3\pi/4)= - \left(\frac{2}{\pi x}\right)^{1/2}\cos (x-\pi/4)
\ee
and taking $b(m_{n})\sim 1$ one obtains 
\be 
f^{(2)}_{n}(\ell)\simeq-\frac{e^{k\ell}}{N_{n}}\left(\frac{2 k}{\pi m_{n} e^{k\ell}}\right)^{1/2} \sqrt{2}(-1)^n.
\ee
Next we may look at $y=0$ in the same mass regime $m_{n}>>k$, where now $x=m_{n}/k$.  Similar manipulations result in 
\be 
f^{(2)}_{n}(0)\simeq -\frac{1}{N_{n}} \left(\frac{2 k}{\pi m_{n}}\right)^{1/2}\sqrt{2}\cos(m_{n}/k).
\ee
Taking $\cos(m_{n}/k)$ to be order $1$ we may define the eigenfunction from the IR brane to UV brane as 
\be
 f^{(2)}_{n}(0)f^{(2)}_{n}(\ell)\simeq\frac{2}{N_{n}} \left(\frac{2 k}{\pi m_{n}}\right)^{1/2} \frac{e^{ k\ell}}{N_{n}}\left(\frac{2 k}{\pi m_{n} e^{k\ell}}\right)^{1/2}(-1)^n
\ee
with 
\be
N_{n}\simeq \frac{1}{\sqrt{m_{n}e^{-k\ell}\pi \ell}}.
\ee
Some simplifications finally result in
\be
 f^{(2)}_{n}(0)f^{(2)}_{n}(\ell)\simeq 4(k\ell)(-1)^n e^{-k\ell/2}.
\ee
An eigenfunction for the derivative of the negative parity fields can be computed similarly,
\be
\partial_5g^{(4)}_{n}(0) \partial_5g^{(4)}_{n}(\ell) \simeq f^{(2)}_{n}(0)f^{(2)}_{n}(\ell) m^2_{n}\simeq 4 m^2_{n}(k\ell)(-1)^n e^{3 k\ell/2}.
\ee
%%%%%%%%%%%%%%%%%%%%%%%%%%%%%%%%%%%%%%%%%%%%%%%%%%%%%%%%%%%%%%%%%%%%%%%%%%%%%%%%%%%%%%%%%%%%%%%%%%%%%%%%%%%%%%%%%%%%%%%%%%%%%%%%%%%%%%%%%%%%%%%%%%%%%%%%%%%%%%%%%%%%%
Now we turn to the regime $m_{n}\ll k$.  These states are highly localise at the IR brane and cannot propagate significantly across the bulk. To see this, we evaluate $f^{(2)}_n(y)$ at $y=0$
\be
 f^{(2)}_{n}(0)\simeq \frac{x}{2N_{m}}\simeq 0,
\ee
where $x=m_{n}/k\ll 1$, for which we used
\be
 J_{1}(x)\simeq x/2 \  , \ \ \ \    Y_{1}(x)\simeq -\frac{2}{\pi x}.
\ee
This is exponentially suppressed and we see that there is very little probability to be located near the UV brane for $m_{n}\ll k$.  Conversely, at y=$\ell$ we use the mass spectrum $m_{n}\simeq (n-1/4)\pi k^*$ and find 
\be
 f^{(2)}_{n}(\ell)\simeq -\frac{e^{ k\ell/2}}{N_{n}}\left(\frac{2 k}{m_{n}\pi}\right)^{1/2}.
\ee
This demonstrates that the light modes are significantly localised to the IR brane not to propagate supersymmetry breaking effects across the interval.  
\subsection{Matsubara frequency summation}
An accurate determination of the full summation of propagators of the Kaluza-Klein tower should be done numerically.  We make a simplifying assumption that we may carry out a Matsubara frequency summation of the KK modes, by approximating the whole KK tower with the states of $m_{n}\gg k$, which have masses $m_{n}\simeq n\pi k^*$.  The resulting summations  give
\be 
\frac{k^*}{2} \sum_{m_{n}}\frac{(-1)^n}{p^2+m^2_{n}}\sim \frac{1}{2}\frac{1}{p \sinh (p/k^*)}
\ee
and
\be 
\frac{k^*}{2} \sum_{m_{n}}\frac{1}{p^2+m^2_{n}}\sim \frac{1}{2} \frac{1}{p \tanh (p/k^*)}
\ee
such that 
\be
\sum_{n}\frac{k^*}{2\ell k^*} \frac{f^{(2)}(0)f^{(2)}(\ell)}{p^2+m^2_{n}}\sim  \frac{2  e^{ k \ell/2 }}{p\sinh (p/k^*)}.
\ee
Similarly, using a contour pulling argument detailed in \cite{Mirabelli:1997aj}, we may convert
\be 
\frac{k^*}{2} \sum_{m_{n}}\frac{1}{p^2+m^2_{n}}\sim \frac{1}{p}\frac{1}{e^{2p/k^*}-1}+ \int^{\infty}_{-\infty} \frac{dp_{5}}{(2\pi)}.
\ee
The second term is independent of $k^*$ and is divergent.  The first term is the finite part that is relevant for computing the Casimir energy.

%\bibliographystyle{JHEP}
%\bibliography{warp.bib}

\providecommand{\href}[2]{#2}\begingroup\raggedright\endgroup

\end{document}